\documentclass[twocolumn,twoside,slac_two]{revtex4}
\usepackage{graphicx}
\usepackage{fancyhdr}
\begin{document}

\title{Precision Charm Decays: Leptonic, Semileptonic, Hadronic}

%

\author{Roy A. Briere}
\affiliation{Carnegie Mellon University, Pittsburgh, Pennsylvania, USA}

\begin{abstract}
Precision results on the leptonic, semileptonic, and hadronic decays 
of charm mesons are reviewed.  
These results are important for timely progress in weak flavor physics 
both in their own right and as checks of theoretical calculations.  
Recent years have seen much progress on weak decays of charm, 
in both theory and experiment, a trend we expect to continue.  
\end{abstract}

\maketitle

\thispagestyle{fancy}

\section{Introduction}

The physics of charm mesons has been long studied for its own interesting 
phenomenology, as the first heavy quark.  But it was, until recently, 
less prized for insight into weak flavor physics than the study of 
$B$ physics, involving the yet heavier but still accessible $b$ quark.  
The higher energy scale afforded $B$ physics a firmer theoretical base, 
and the surprising suppression of the leading $b \to  c W^-$ decay process 
allows for strong displays of $CP$ violation and neutral meson mixing.  
In the case of charm, less interesting Cabibbo-allowed decays dominate, 
making it hard to see rare processes.   
Charm $CP$ violation is generally suppressed in the Standard Model, 
as is $D^0$ mixing, which is also complicated by uncertain long-distance 
contributions.  
Experimental results are in general harder to interpret since charm 
is light, making calculations of QCD effects difficult.  
The ``brown muck'' of strong-interaction physics obscures 
the weak physics of interest.  
Of course, not all was bleak; there were opportunities beyond 
mere cataloging of charm decays.   
One looked for unexpected surprises, or exploited the benefits of using 
charm to limit new physics in the light of small Standard Model backgrounds.  
However, a new communal wisdom began to emerge not long ago, 
that charm is a gift with many uses \cite{Cicerone}.  
 
Much of the activity in weak flavor physics involves 
precision measurements to over-constrain the Cabibbo-Kobayashi-Maskawa 
(CKM) quark mixing matrix.  An inconsistency would signal new physics, 
and searching for such indirect evidence complements direct searches 
for new particles at the energy frontier.  
$B$ physics experiments are the most fertile ground for these studies; 
they are very productive and precise, but too often limited by theory.  
CKM constraints are often shown as bands in a two-dimensional 
($\rho-\eta$) plane, and the widths of many of these constraint bands 
are limited by theory.  
Lattice QCD (LQCD) provides a systematic path to lift such limitations 
in using $B$ results, and charm physics can be used to directly test LQCD.  
Of course, this is not all that charm has to offer.  
Recently, multiple consistent observations of $D^0$ mixing 
have enlivened this topic, as only discovery can.  
And analyses of charm meson Dalitz plots provide a useful tool 
not only for low-energy hadronic spectroscopy, but are in fact 
also helpful for improving precision of weak $B$ physics results, 
in particular for the CKM angle $\gamma$.  

Here, we concentrate on precision branching fractions results 
which impact weak flavor physics, though we discuss some other 
interesting hadronic modes at the end.  
In particular, the increased precision of leptonic and 
semi-leptonic results provides a testbed for modern LQCD.  
If it passes, we add confidence in LQCD 
predictions and uncertainty estimates for $B$ physics, where they 
are a critically needed compliment to pre-existing measurements 
of $B$ mixing and exclusive $B$ semileptonic decays.  

Leptonic decays of $D_{(s)}$ are used to extract decay constants 
and the semileptonic processes $D \to K\ell\nu, \pi\ell\nu$ constrain 
form-factors.  Golden-mode branching ratios for 
$D^0 \to K^-\pi^+$, $D^+ \to K^-\pi^+\pi^+$, $D_s \to K^-K^+\pi^+$ 
solidify the overall normalization of charm decays.  
Precision lifetimes, now dominated by FOCUS \cite{PDG2008} results, are also 
an important input to connect theoretical partial widths to 
expected branching fractions.  
Relevant experimental techniques include tagging with near-threshold 
$D$ pairs at CLEO-c and high-statistics 10 GeV data from 
$B$ factories which allows for what we call ``continuum tagging''.  

\section{Experiments and Techniques}

Many results have been obtained with $e^+ e^-$ collisions 
at threshold by CLEO-c.  
For $D$ mesons, machine energies are set to produce 
$e^+ e^-\to \psi(3770) \to D^0\bar{D}^0, D^+D^-$, 
while for $D_s$, one uses $e^+ e^-$ collision at 4170 MeV 
to produce $D_s^+ D_s^{*-} + c.c.$ 
At the $\psi(3770)$, the $D$-pair cross-section is 
$\sigma(D\bar{D}) = (6.57\pm 0.04 \pm 0.10)$ nb \cite{cleo_dhad}, 
while at 4170 MeV, $\sigma(D_s^+ D_s^{*-} + c.c) 
= (0.916 \pm 0.011 \pm 0.049)$ nb \cite{cleo_scan}.  
These desired processes are in addition to a light-quark ($uds$) 
continuum cross-section of order 20 nb, plus tau pairs, radiative 
returns to the narrow $J/\psi$ and $\psi'$, and two-photon reactions.  
But note that charm mesons at 3770 MeV appear {\it only} 
as $D\bar{D}$ pairs, with no room for even one fragmentation pion.  
On the other hand, data at 4170 MeV also includes 
$D\bar{D}, D^*\bar{D} + D\bar{D}^*, D^*\bar{D^*}$, 
and a small amount of $D_s^+D_s^-$ in addition to the desired 
$D_s^+ D_s^{*-}$ (which is chosen due to a larger cross-section 
relative to the simpler $D_s^+ D_s^-$).  
Kinematics can still cleanly separate these different possibilities 
at 4170 MeV, but the other charm and lower cross-section for the 
process of interest does lead to more combinatorics and decreased 
signal-to-noise in the $D_s$ case.  

A ``tag'' is simply a fully-reconstructed $D_{(s)}$ hadronic decay.  
A sample of tagged events has greatly reduced background 
and constrained kinematics, both of which aid studies 
of how the other $D_{(s)}$ in the event decays.  
One can infer neutrinos from energy and momentum conservation, 
allowing ``full'' reconstruction of (semi)leptonic decays.  
For hadronic branching fractions, some simple algebra will serve to 
demonstrate how tagged samples can provide a clean absolute normalization.  
The typical tag rates per $D_{(s)}$ ({\it not} per pair) are roughly 
15\%, 10\%, and 5\% for $D^0, D^+$, and $D_s$, respectively.  
CLEO-c completed data-taking in Spring, 2008, and is in the 
process of updating many results to the full statistics.  
In the meantime, the BESIII experiment at the BEPCII collider 
has begun its commissioning run, and their plans are 
to extend these methods even further.  

Belle has developed tag-based techniques for use with continuum charm 
production from the $\Upsilon$ region, near 10 GeV.  With hundreds 
of times greater charm samples, one can afford to use selected favorable 
semi-exclusive states which are only a small fraction of the total.  
This technique is exploited by Belle for (semi)leptonic decays.  
Partial reconstruction techniques may be used to study charm produced 
in $B$ decays, a production mechanism with a rate similar to the 
$c\bar{c}$ continuum.  
A precision branching fraction result from BaBar is featured 
to illustrate these techniques.  
We will see that $D^*$ tagging of $D$ mesons is no longer the main workhorse.  

\section{Leptonic Decays and Decay Constants}

Decay constants characterize the strong-interaction physics 
at the quark-annihilation vertex, in our case $c\bar{d},c\bar{s} \to W^+$.  
In a fully leptonic decay, they parameterize all of our essential 
theoretical limitations.  
Decay constants also appear in the evaluation of box diagrams, 
and limit theoretical precision in calculating the neutral meson mixing.  
Thus, lack of knowledge of the $B^0$ and $B_s$ decays constants 
limits the usefulness of precise measurements of $B^0-\bar{B}^0$ and 
$B_s-\bar{B}_s$ oscillations.  These mixing data are our best source 
of information on the CKM matrix elements $V_{td}$ and $V_{ts}$, 
which are difficult to measure directly in top decay.    
The charm sector presents an opportunity to check LQCD results 
for decay constants against precision measurements.  
This is desirable since the only related direct measurement 
possible, the decay constant of the $B^+$ from measurement of 
$B^+ \to \ell^+\nu$, is very challenging.  

Let us first review the status in the Spring of 2008.  
Previous CLEO and Belle results for $D_s$ leptonic decays averaged 
to give $f_{D_s} = (273 \pm 10)$ MeV \cite{RosnerStone}.  
The best precision claimed for LQCD, with 2+1 unquenched flavors, 
finds $(241 \pm 3)$ MeV \cite{LQCD_Follana}, about a three 
standard-deviation discrepancy.  
On the other hand, $D^+$ results are consistent with LQCD.  
A 2005 CLEO $f_{D}$ result of $f_D = 223 \pm 17$ MeV 
from 281 pb$^{-1}$ \cite{CLEO_old_fd}, is consistent 
with a recent LQCD value of $(207 \pm 4)$ MeV \cite{LQCD_Follana}.  
However, in both cases, more experimental precision is clearly needed.  
We also await more LQCD results, to confirm the quoted results.  
While these results have the best claimed precision, they involve 
a somewhat controversial technique involving a ``fourth-root trick'' 
using computationally-efficient staggered fermions.  
Further details and other competing lattice results may be found in Refs. 
\cite{RosnerStone} and \cite{Lattice2008}.  
Most, but not all, have similar central values; they have larger errors 
due primarily to computational limitations with more traditional 
(but also more generally accepted) LQCD techniques.  

Dobrescu and Kronfeld \cite{DK_fdtheory} argue that the disagreement 
on $f_{D_S}$ could be the effect of new physics, mentioning 
both charged Higgs (their own model) and leptoquarks as possibilities.  
Kundu and Nandi suggest \cite{KN_fdtheory} $R$-parity violating SUSY 
to explain a large $f_{D_s}$ and the $B_s$ mixing phase.

\begin{figure}
\includegraphics[width=65mm]{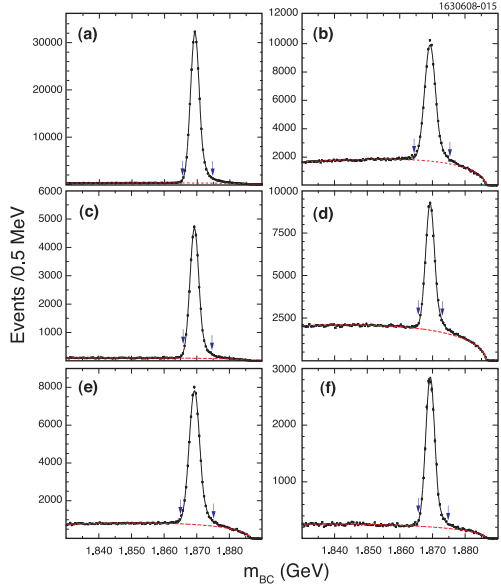}%
\caption{\label{fig:CLEO_fdtags} 
   CLEO-c $D^+$ hadronic tags used for the $D^+ \to \mu^+\nu$ analysis. 
   Modes in panels (a)$-$(f) are 
   $K^+\pi^-\pi^-$, $K^+\pi^-\pi^-\pi^0$, $K_S\pi^-$, 
   $K_S\pi^-\pi^-\pi^+$, $K_S\pi^-\pi^0$, and $K^+K^-\pi^-$, respectively.}
\end{figure}

\begin{figure}
\includegraphics[width=50mm]{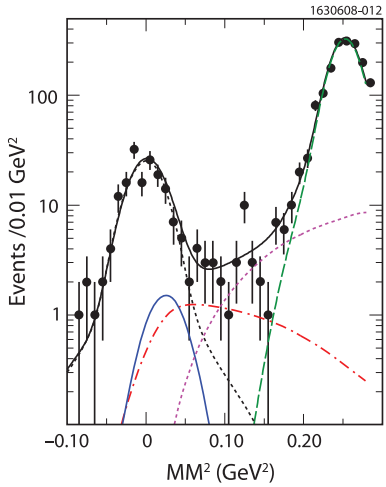}%
\caption{\label{fig:CLEO_fd} 
   CLEO-c missing-mass-squared plot for $D^+ \to \mu^+\nu$; 
   note the vertical log scale.  The solid curve is a fit to the data points.  
   Components of the fit include $\mu\nu$ signal (short-dash black), 
   $\tau\nu$ (dot-dash red), residual $\pi^+\pi^0$ background (solid blue), 
   $K\pi$ background (long-dash green), 
   and other backgrounds (short-dash purple).}
\end{figure}

There are now newer experimental results to report on.  
We first present a preliminary CLEO-c $D^+\to \mu^+ \nu$ update 
using the full 818 pb$^{-1}$ of $\psi(3770)$ data \cite{CLEO_fd}.  
The signal side opposite the $D^-$ tag is a single track 
and an unobserved neutrino.  
With the neutrino ``detected'' by four-momentum balance, one 
has a fully-reconstructed double-tag event: all decays products 
of both the $D^+$ and $D^-$ are measured.  Mass peaks for 
the six hadronic tag modes are shown in Fig. \ref{fig:CLEO_fdtags}.  
The chosen signal variable for the $\mu^+\nu$ decay is the calculated 
square of the missing-mass of any undetected decay products, shown in 
Fig. \ref{fig:CLEO_fd}; this should of course peak at $M_\nu^2 = 0$.  
The power of $D$-tagging is evident in the clean, isolated signal peak.  
The signal fit includes $K^0 \pi^+$, $\pi^+\pi^0$, $\tau^+\nu$ components 
as well.  Fixing the $\tau$-to-$\mu$ ratio at the Standard Model ratio 
of 2.65 gives 
${\cal B}(D^+ \to \mu^+ \nu) = (3.82 \pm 0.32 \pm 0.09) \times 10^{-4}$ 
and $f_{D} = (205.8 \pm 8.5 \pm 2.5)$ MeV, 
which is the best number in the context of the SM.  
Floating this ratio gives a consistent result of 
${\cal B}(D^+ \to \mu^+ \nu) = (3.93 \pm 0.35 \pm 0.09) \times 10^{-4}$ \
and $f_{D} = (207.6 \pm 9.3 \pm 2.5)$ MeV; 
this is the best number for use with new-physics models.  
One sees that, in either case, this full-statistics CLEO-c result 
is almost twice as precise as their previous one, 
and remains consistent with LQCD.  

\begin{figure}
\includegraphics[width=75mm]{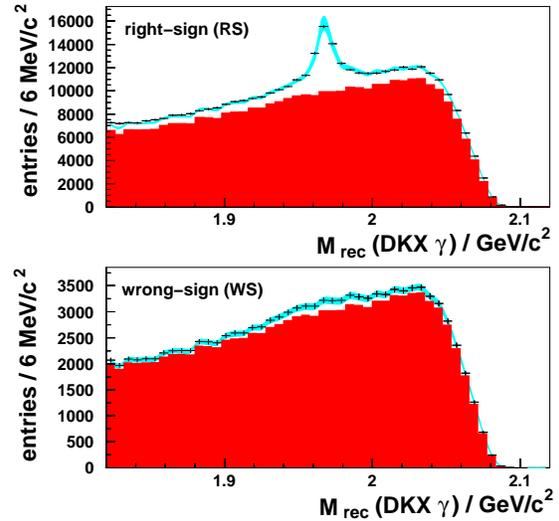}%
\caption{\label{fig:belle_fds1} 
    Recoil mass against the $DKX\gamma$ system from the 
    Belle $D_s \to \mu\nu$ analysis.  
    Both right-sign (RS) and wrong-sign (WS) are shown.  
    Dark red shading shows the fitted background, while 
    the light blue shading shows the total fit to signal 
    and background with systematic uncertainties.}
\end{figure}

\begin{figure}
\includegraphics[width=75mm]{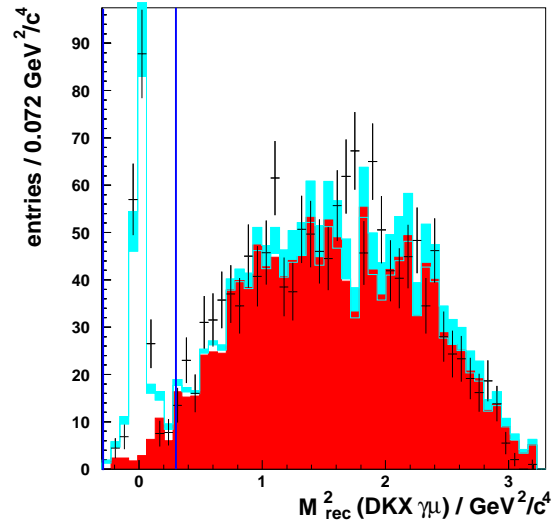}%
\caption{\label{fig:belle_fds2} 
     Recoil mass squared against the $DKX\gamma\mu$ system from the 
     Belle $D_s \to \mu\nu$ analysis.  
     The color scheme is the same as the previous plot.  
     The signal region is indicated by vertical lines near zero.
     }
\end{figure}

Now we turn to $f_{D_s}$.  While the Belle $D_s^+ \to \mu^+ \nu$ result 
from 548 fb$^{-1}$ \cite{Belle_fds} has been public for some time, 
it is useful to illustrate their technique.  
They use ``continuum tagging'' with the process 
$e^+e^-\to D^{\pm,0} K^{\pm,0} X D_s^{*-}$, 
where $X = n\pi$ or $n\pi \gamma$ (from fragmentation).  
About 25\% of the $D$ decay modes are used to reconstruct 
the tag side.  
The branching ratio of $D_s \to \mu\nu$ is proportional to a ratio 
of two yields, both obtained from recoil mass peaks.  
The recoil against $DKX\gamma$ peaks at the $D_s$ mass, and counts the 
total number of $D_s$ in their sample; see Fig. \ref{fig:belle_fds1}.  
The recoil against $DKX\gamma\mu$ peaks at 0 ($m_\nu^2$), and is used 
to count $D_s \to \mu \nu$ decays; see Fig. \ref{fig:belle_fds2}.  
After some sophisticated fitting, they obtain a final result of: 
${\cal B}(D_s^+ \to \mu^+\nu) = (0.644 \pm 0.076 \pm 0.057)\%$ 
and $f_{D_s} = (275 \pm 16 \pm 12)$ MeV.  

\begin{figure}
\includegraphics[width=70mm]{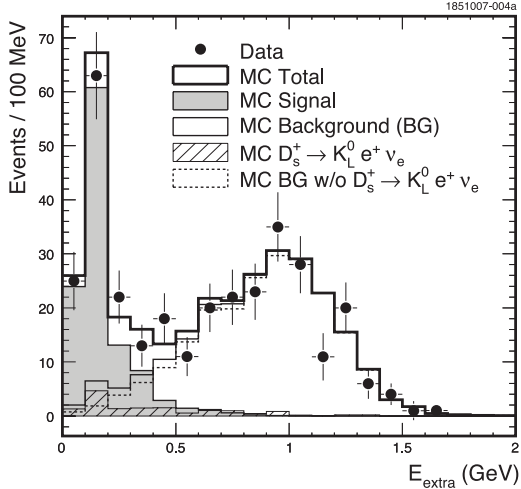}%
\caption{\label{fig:CLEO_fDs_e} 
    A plot of $E_{extra}$ for the CLEO-c $D_s^+ \to \tau^+\nu$, 
    with $\tau^+ \to e^+\nu\bar{\nu}$ analysis.  
    $E_{extra}$ is the energy in the calorimeter not 
    associated with the $D_s$ tag or the electron.  
    Semileptonic backgrounds tend to peak away from zero.  
    The signal peak includes cases where the $140$ MeV 
    transition photon from $D_s^* \to D_s \gamma$ 
    is included in $E_{extra}$, as well as cases where it 
    is missed.  }
\end{figure}

CLEO-c has published an analysis of $D_s^+ \to \tau^+\nu$, 
with $\tau^+ \to e^+\nu\bar{\nu}$ using 298 pb$^{-1}$ \cite{CLEO_fds_e}.  
An update to the full statistics with an improved analysis is in progress.  
For their published analysis, only the three cleanest $D_s^-$ tags 
($\phi\pi^-, K^-K^{*0}, K^-K_S$) are used.  
There is always more than one neutrino, making the kinematics 
less well-constrained than the $\mu^+\nu$ final state.  
The amount of extra energy (not from the $D_s^-$ tag or the $e^+$) 
in the calorimeter, $E_{extra}$, is used as the main signal variable;  
see Fig. \ref{fig:CLEO_fDs_e}.  
The signal region defined as $E_{extra} < 400$ MeV.  
Most background is from semileptonic events, which tend to have 
some extra energy from their hadronic daughters.  
Care must be taken with $D^+ \to K_L e^+ \nu$ background which peaks 
lower than other semileptonic modes.  
Note that a reconstructed $\gamma$ from the $D_s^* \to D_s \gamma$ 
decay is not required.  
The final result is: 
${\cal B}(D_s^+ \to \tau^+\nu) = (6.17 \pm 0.71 \pm 0.34)\%$, 
and $f_{D_s} = (274 \pm 16 \pm 8)$ MeV.  

\begin{figure}
\includegraphics[width=70mm]{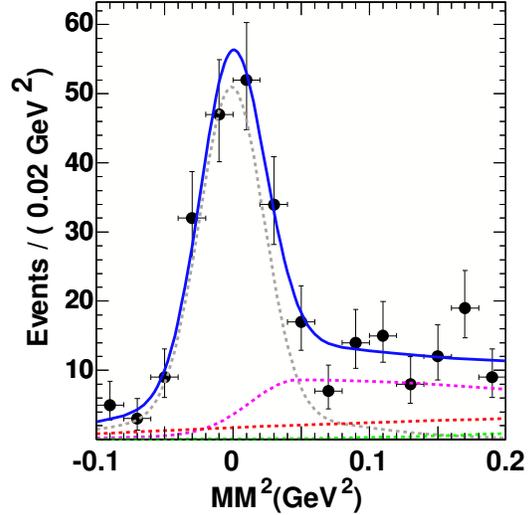}%
\caption{\label{fig:CLEO_fds} 
    Missing-mass-squared plot for the CLEO-c $D_s \to \mu\nu, \tau\nu$ 
    with $\tau \to \pi\nu$ analysis.  
    The solid blue curve is a fit to the data points.  
    Fit components include $\mu\nu$ signal (dashed grey), 
    $\tau\nu$ signal (dashed purple), and two small background 
    components (dashed green and red).}
\end{figure}

The most precise CLEO-c results come from an analysis of $D_s^+ \to \mu^+\nu$ 
and $D_s^+ \to \tau^+\nu, \tau^+ \to \pi^+\nu$.  They are analyzed together 
since both contain a single track from the signal (non-tag) $D_s$.  
Note that explicit muon identification is unnecessary and would actually 
be inefficient at the ${\cal O}$(1 GeV) energies relevant here.  
A published result using 314 pb$^{-1}$ and eight hadronic decay modes 
is available \cite{CLEO_old_fds}; we present here a preliminary update 
using $\sim 400$ pb$^{-1}$ \cite{CLEO_fds}.  
As with the $D^+$ analysis, the signal is studied via the 
missing-mass-squared; see Fig. \ref{fig:CLEO_fds}.  
With the $\mu\nu$ to $\tau\nu$ ratio fixed at the Standard Model value 
in the signal fit they find: 
${\cal B}^{\rm eff} (D_s^+ \to \tau^+ \nu) = (0.613 \pm 0.044 \pm 0.020)$, 
and $f_{D_s} = (268.2 \pm 9.6 \pm 4.4)$ MeV.  
Fitting with both floating yields consistent results, and gives 
$\Gamma(D_s^+ \to \tau\nu)/\Gamma(D_s^+ \to \mu^+\nu) = 10.3 \pm 1.1$.  
This is consistent with the Standard Model expectation of 9.72, 
and hence limits some new physics scenarios.  

A new preliminary CLEO-c average of this $\mu\nu$ plus $\tau\nu, 
\tau \to \pi\nu$ with the previous $\tau \nu, \tau \to e\nu\nu$ result 
yields: $f_{D_s} = (269.4 \pm 8.2 \pm 3.9)$ MeV.  A weighted average 
of CLEO-c and Belle gives $f_{D_s} = (269.6 \pm 8.3)$ MeV.  
We summarize the current situation for $f_{D_s}$ in Table \ref{tab:fDs}.  
Note that a BaBar result on $f_{D_s}$ \cite{BaBar_fds} and most other older 
analyses depend on ${\cal B}(D_s \to \phi\pi)$, and are omitted here.  
\begin{table*}[t]
\begin{center}
\caption{Summary of recent experimental results on $f_{D_s}$.}
\begin{tabular}{|l|c|c|c|}
\hline
Experiment & Luminosity & mode & $f_{D_s}$ \\
\hline 
CLEO-c \cite{CLEO_fds} (prelim) & $\simeq 400$ pb$^{-1}$ & 
   $D_s \to \mu\nu, \tau\nu(\tau \to \pi\nu)$ & 
   $268.2 \pm 9.6 \pm 4.4$ \\
CLEO-c \cite{CLEO_fds_e}& 298 pb$^{-1}$ & 
   $D_s \to \tau\nu (\tau \to e\nu\nu)$ & $273 \pm 16 \pm 8$ \\
CLEO-c Average  \cite{CLEO_fds} (prelim) & - & - & 
   $267.9 \pm 8.2 \pm 3.9$ \\
\hline
Belle \cite{Belle_fds} & 548 fb$^{-1}$ & $D_s \to \mu\nu$ &
   $275 \pm 16 \pm 12$ \\
\hline
Current CLEO-c/Belle Average \cite{CLEO_fds} & - & - & 
   $269.6 \pm 8.3$ \\
\hline
\end{tabular}
\label{tab:fDs}
\end{center}
\end{table*}
The roughly three standard deviation disagreement between theory and 
experiment for $f_{D_s}$ still remains.  
Is it a problem with the theory, or experiment? 
A hint of new physics?  Or perhaps partly an unlikely fluctuation?  
One must also keep in mind the excellent agreement obtained for $f_D$.

\section{Semileptonic Decays and Form Factors}

Here, strong-interaction complications are 
summarized by form factors, which are functions of $q^2 = M^2_{\ell\nu}$.  
Once again, charm offers a chance to confront LQCD calculations.  
Similar LQCD calculations can help us to interpret results on 
$B \to \pi\ell\nu$ in order to extract $V_{ub}$.  
The key charm mode for tests is $D^0 \to \pi^-\ell^+\nu$.  
We discuss this and related pseudo-scalar modes; 
later, a new precision result on $D_s^+ \to K^+K^-e^+\nu$ is also featured.  
Form-factor normalizations can be used with external $V_{cs}, V_{cd}$ as LQCD
tests, or utilizing LQCD, for extractions of $V_{cs}$, and $V_{cd}$.  
We could also use ratios with leptonic decays to cancel these CKM matrix 
elements and test the ratio of LQCD predictions for the two types of decays.  
The form-factor shapes as a function of $q^2$ also provide direct LQCD tests.  
We note that $d\Gamma/dq^2$ drops rapidly due to a $p^3$ phase-space factor, 
decreasing statistics and precision at large $q^2$.  

Belle used 282 fb$^{-1}$ to study $D^0 \to \pi^- \ell^+\nu, K^-\ell^+\nu$ 
\cite{Belle_semilep}.   
They use ``continuum tagging'' again here, exploiting 
$e^+e^- \to \bar{D}^{(*)}_{tag} D^*_{signal} X$.  
They reconstruct all particles, except for neutrino, and use 
$m^2_{miss} = E^2_{miss} - p^2_{miss}$ as their signal variable.  
Tagging provides absolute normalization of about 56,000 tagged $D^0$.  
They find $126 \pm 12$ $ D^0 \to \pi^-e^+\nu$ events, and   
$106 \pm 12$ $D^0 \to \pi^-\mu^+\nu$ events.  

CLEO-c has a published untagged analysis which uses global four-momentum
balance to infer the neutrino without an explicit $D$ tag 
\cite{CLEO_sluntag}.  This has about twice the efficiency 
of the CLEO-c simpler tagged analysis discussed next.  
Fitting the familiar beam-energy-constrained mass, they find 
$1325 \pm 48$ $D^0 \to \pi^-e^+\nu$ events.  

\begin{figure}
\includegraphics[width=70mm]{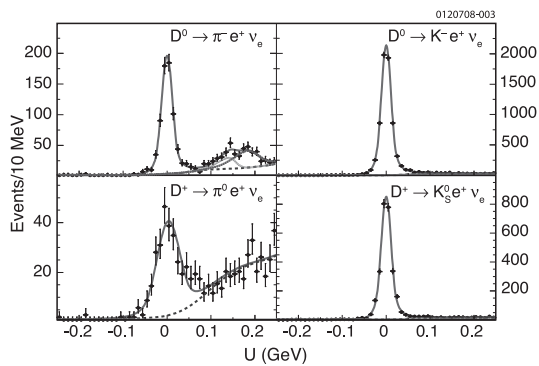}
\caption{\label{fig:CLEO_sltag1} 
    CLEO-c tagged semileptonic analysis signal peaks in 
    $U = E_{miss} - p_{miss}$ for the four $K e \nu, \pi e \nu$ modes.  
    Notice how clean the data is, despite the neutrino. }
\end{figure}

\begin{figure}
\includegraphics[width=70mm]{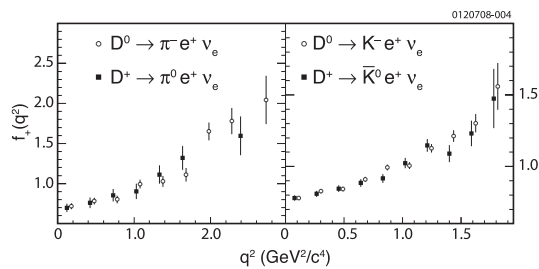}%
\caption{\label{fig:CLEO_sltag2} 
    Form factors extracted from the CLEO-c tagged semileptonic analysis.
    Isospin-related modes are plotted together and agree well, as expected.}
\end{figure}

There is also a CLEO-c preliminary result also using 281 pb$^{-1}$, 
but with $D$ tagging \cite{CLEO_sltag}.  
Signal peaks in are shown in Fig. \ref{fig:CLEO_sltag1} 
and the extracted form factors in Fig. \ref{fig:CLEO_sltag2}.  
All four $K e \nu, \pi e \nu$ modes are studied, 
using $U_{miss} = E_{miss} - |p_{miss}|$ as the signal variable; 
see Fig. \ref{fig:CLEO_sltag1}.  
They find $699 \pm 28$ $D^0 \to \pi^-e^+\nu$ events; 
threshold kinematics and tagging result in very small contamination 
from the ten-times more copious $K^- \pi^+\nu$ mode.  
Their paper also discusses correlations and performs averages 
with the untagged analysis.  

A more traditional $D^0 \to K^-\ell^+\nu$ analysis was done at BaBar 
by tagging with $D^{*+} \to D^0\pi^+$ \cite{Babar_kenu}.  
While yielding very large signal statistics and an excellent analysis 
of $K^+\ell^-\nu$ form factors, 
this technique is not well-suited to the rarer $\pi^-\ell^+\nu$ mode.  
It also has somewhat poorer $q^2$ resolution, with unfolding needed 
for the form factor measurements.  

All together, significant improvements in precision have been obtained 
by these recent BaBar, Belle, and CLEO-c measurements.  
CLEO-c is the most precise for $\pi^- \ell^+ \nu$, obtaining about 
three times more signal events and ten times the signal-to-noise 
in the tagged case.  
Both experiments have of order three times more luminosity to analyze.  
Note that Belle uses electrons and muons, while CLEO-c uses only electrons.  
However, CLEO also does the two isospin-related modes with neutral hadrons.  

We will not discuss detailed comparisons with LQCD here; 
some comparisons can be found in the references.  
Agreement is good, within current errors.  
Currently, normalization errors are about 10\% for theory for both $Ke\nu$ 
and $\pi e \nu$, and  are 2\% and 4\%, respectively, from experiment.  
For $Ke\nu$, a single pole model fit works reasonably well, 
but not with the spectroscopic $D_s^*$ mass as the pole mass.  
One key issue is that fitting to a model that is not a correct description, 
such as a simple pole model, leads to extracted results that cannot be
sensibly compared, since data points from different sources, 
experiments and LQCD alike, will have different precisions vs. $q^2$.  
An incorrect theory fit will be biased by preferentially accommodating 
the more precise points being fit.  
A recent influential theory paper \cite{BecherHill} advocates using 
well-motivated series expansions to avoid such problems.  
The newer analyses (from BaBar and CLEO-c) use these expansions along 
with some older pole forms.  

We finish with a new preliminary BaBar result on $D_s^+ \to K^+K^- e^+\nu$ 
form-factors based on 214 fb$^{-1}$ \cite{BaBar_kkenu}.  
For this Cabibbo-allowed mode, high-energy continuum charm data give 
high statistics and are preferable to charm-threshold data.  
Their reconstruction technique exploits the jet structure to approximate 
some of the kinematics and then fits to constrain the neutrino.  
A detailed form-factor analysis obtains excellent angular fits 
to their high-statistics sample of about 25,000 events.  
While dominated by $D_s \to \phi e \nu$, they also find the 
first evidence for $D_s \to f_0 e \nu$.  
There is one form-factor for $f_0 e \nu$; 
their fit floats the intercept at $q^2 = 0$ and the pole mass $m_A$.  
The vector $\phi$ meson is more complicated, requiring three form-factors 
for $\phi e \nu$.  Their fit floats two relative normalizations 
using a fixed $m_V$ as a common pole for the shape.  
Overall, good agreement with LQCD \cite{LQCD_KKenu} is obtained, 
except perhaps for one of the normalizations in the $\phi e\nu$ case.  

Some other interesting results have fallen victim to space constraints, 
notably inclusive branching ratios \cite{CLEO_incl} 
as well as other exclusive branching ratios, 
for $D \to (\rho/\omega/\eta/K_1) e \nu$ \cite{CLEO_other_sl}.

\section{Precision Hadronic Branching Fractions}

We begin by explaining the tagging method used at CLEO-c.  
The number of ``single tags'' observed where a $D$ decays in mode $j$ is 
$N_j = N_{D\bar{D}} B_j \epsilon_j$; where $N_{D\bar{D}}$ is the number of 
$D\bar{D}$ pairs created, and ${\cal B}$ and $\epsilon$ are the 
branching fraction and efficiency.  
Similarly, the number of ``double tags'', where 
the $D$ decays via mode $i$ and the $\bar{D}$ via mode $j$ is 
$N_{ij} = N_{D\bar{D}} B_i B_j \epsilon_{ij}$.  
Simple algebra then gives 
$B_i = (N_{ij} \epsilon_j) / (N_j \epsilon_{ij})$ and  
$N_{D\bar{D}} = (N_i N_j \epsilon_{ij}) / (N_{ij} \epsilon_i \epsilon_j)$.  
Two key advantages are evident here.  
First, ${\cal B}_i$ independent of $N_{D\bar{D}}$ and the integrated luminosity; 
absolute normalization here essentially comes from the algebra.  
Second, one can measure cross-sections independent of branching fractions.  
We also note that $\epsilon_j / \epsilon_{ij} \simeq \epsilon_i$, making 
the branching fraction measurement almost independent of the tag mode $j$.  
This is true since we have low-multiplicity decays in a fine-grained
detector.  Monte-Carlo simulations are used for the signal-mode 
efficiency and the small effects of tag mode (the efficiency approximation 
noted above is {\it not} assumed).  
Systematics uncertainties are dominated by charged track, $\pi^0$, $K_S$, 
and particle identification efficiencies.  
It is important that $D$ tagging techniques can also be used to explore 
the absolute efficiency scale with the same data sample \cite{cleo_dhad}.  
Three $D^0$ and six $D^+$ modes are used, and results are obtained 
from a global fit to all single-tag and double-tag rates with correlations 
properly included.  
Their results include: 
${\cal B}(D^0 \to K^- \pi^+) 
= ( 3.891 \pm 0.035 \pm 0.059 \pm 0.035 )\%$ and 
${\cal B}(D^+ \to K^- \pi^+ \pi^+) 
= ( 9.14 \pm 0.10 \pm 0.16 \pm 0.07 )\%$.  
Uncertainties are statistical, systematic (dominated by knowledge of 
absolute efficiencies), and effects of final-state radiation.  

\begin{figure}
\includegraphics[width=70mm]{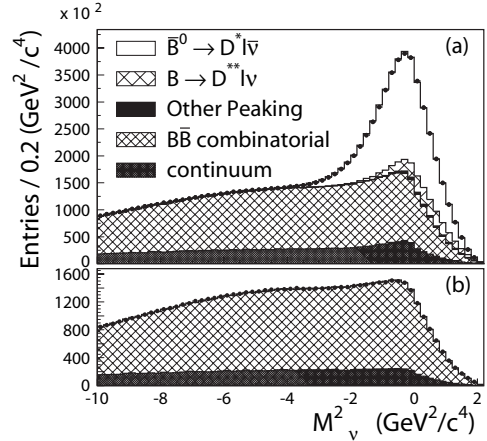}%
\caption{\label{fig:BaBar_kpi1} 
     The missing-mass distribution from the BaBar $D^0 \to K^-\pi^+$ 
     analysis, showing the peak from the desired 
     $\bar{B}^0 \to D^{*+} \ell^- \nu$ process.  
     This peak provides the normalization for the $D^0 \to K^-\pi^+$ 
     branching ratios.  
     Wrong-sign $D^* \ell$ pairs are also shown.}
\end{figure}

\begin{figure}
\includegraphics[width=70mm]{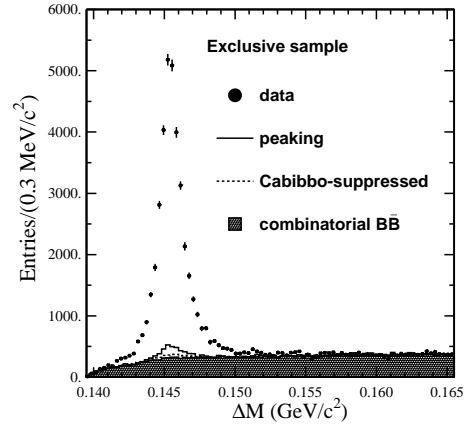}%
\caption{\label{fig:BaBar_kpi2} 
    The BaBar $D^0 \to K^-\pi^+$ signal, shown as a peak in 
    the $D^{*+} - D^0$ mass difference, $\Delta M$. }
\end{figure}

BaBar has also measured $D^0 \to K^- \pi^+$ with 232 fb$^{-1}$ 
of data at or near the $\Upsilon(4S)$ \cite{babar_brkpi}.  
They employ a partial reconstruction technique for the decay 
$B^0 \to D^{*+} (X) l \nu$.  The slow pion from the $D^{*+} \to D^0 \pi^+_{slow}$ 
decay is used to estimate the $D^{*+}$ momentum and calculate the missing 
mass without explicit reconstruction of the $D^0$.  
The resulting missing-mass peak counts the number of $D^0$ mesons in this 
inclusive sample; see Fig. \ref{fig:BaBar_kpi1}.  
Full reconstruction of $D^0 \to K^- \pi^+$ is then performed 
within this inclusive sample to count the $D^0 \to K^-\pi^+$ decays; 
see Fig. \ref{fig:BaBar_kpi2}.  
${\cal B}(D^0 \to K^- \pi^+)  = ( 4.007 \pm 0.037 \pm 0.072 )\%$.  
Systematics of 1.8\% primarily come from a 1.5\% uncertainty on the 
exclusive efficiency and 1.0\% for the inclusive efficiency.  

The golden $D^+$ mode is now measured to 2.2\% by CLEO-c, 
the best previous single measurements had errors larger than 10\%.  
The golden $D^0$ mode is now determined to about 2\% by 
both CLEO-c and BaBar; the best previous measurements had 
errors greater than 3.6\%.  
Both of these key modes are now systematics limited.  

\begin{figure}
\includegraphics[width=70mm]{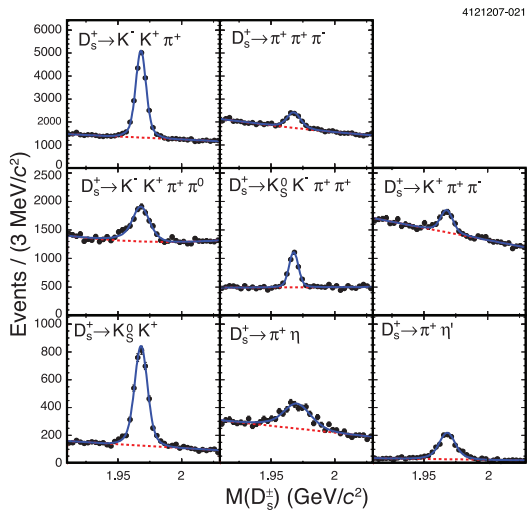}%
\caption{\label{fig:CLEO_Ds} 
         CLEO-c $D_s^+$ mass peaks for hadronic single-tag modes 
         used in the branching fraction analysis.}
\end{figure}

\begin{figure}
\includegraphics[width=70mm]{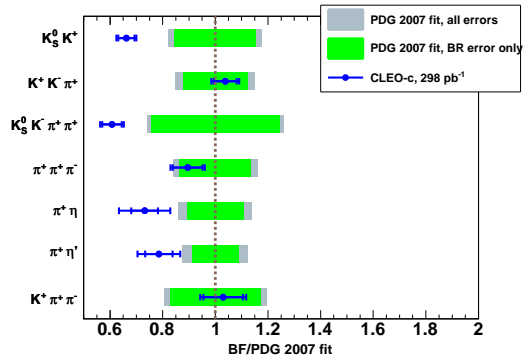}%
\caption{\label{fig:dshad} 
         Summary of CLEO-c $D_s^+$ branching ratio 
         results compared to previous world averages.  }
\end{figure}

More recently, CLEO-c reported hadronic branching fractions 
for several $D_s$ decays, based on 298 pb$^{-1}$ of data 
at $E_{cm} = 4170$ MeV \cite{cleo_dshad}.  
Single-tag mass peaks for the eight $D_s$ decay studied are shown in 
Fig. \ref{fig:CLEO_Ds}.  
Their sample of about $N_{DT} = 1000$ double tags sets the scale 
of achievable statistical uncertainties as $\ge 1/\sqrt{N_{DT}}$.  
We point out that the historical use of $D_s^+ \to \phi\pi^+$ 
as the mode of choice for $D_s$ normalization has some problems.  
It is difficult to cleanly define what ``$\phi$'' means given other nearby 
resonances (such as the $f_0$) and interference between various contributions 
to the $K^+K^-\pi^+$ final state.  These effects change the angular
distributions in the $\phi$ mass region, complicate separation of 
$\phi$ and non-$\phi$ contributions, and lead to dependence on experimental 
resolution.  
Thus, CLEO-c quotes a result for the entire phase-space: 
${\cal B}(D_s \to K^+ K^- \pi^+) = (5.50 \pm 0.23 \pm 0.16)\%$.  
They also quote ${\cal B}(D_s \to K^+ K^- \pi^+)$ 
within various $K^+ K^-$ mass windows, which are related 
to some sort of ``$\phi\pi^+$'' branching ratio.  
In Fig. \ref{fig:dshad}, we illustrate the improvements over 
previous world averages achieved by these new results.  
The full data sample will more than double the statistics.  

While we have concentrated here on golden-mode branching fractions for 
$D^0$, $D^+$ and $D_s$, many other results such 
as Cabibbo-suppressed final states and inclusive branching fractions 
have also been explored \cite{PDG2008}.

\section{Other Hadronic Results}

We start with some of the unique opportunities available at 
CLEO-c due to quantum correlations between the $D\bar{D}$ pairs 
produced at the $\psi(3770)$.  
This is used, for example, to measure the strong $K\pi$ final-state 
interaction (FSI) phase with 281 pb$^{-1}$ \cite{cleo_tqca}.  
This is of great interest for interpreting $D$ mixing results 
based on the $D^0 \to K^-\pi^+$ mode, which are contaminated 
by this FSI phase.  
Results are extracted from a simultaneous fit to their data on 
many hadronic (both flavor and $CP$ eigenstates) and semileptonic modes, 
plus external inputs from $D$ mixing measurements.  
For the experts, these external inputs are $x, y, x', y', r^2$.  
They find: $\cos \delta = 1.10 \pm 0.35 \pm 0.07 $, 
$\delta = (22^{+11}_{-12}$$^{+9}_{-11})^\circ$.  
Note that the most likley $\cos\delta$ is in general {\it not} 
the cosine of the most likely $\delta$, due to a $d\cos\delta/d\delta$ 
factor when changing the abscissa.  

CLEO-c has also investigated interference in $K_L \pi, K_S \pi$ final 
states.  $D$ decay diagrams can produce both $K^0$ and $\bar{K}^0$; 
these interfere in physical $K_L$, $K_S$ final states, 
which leads to a $K_S$, $K_L$ asymmetry \cite{BY}.   
For $D^0 \to K_{S,L} \pi^0$, we expect an asymmetry of 
$R(D^0) = 2 \tan^2\theta_C \sim 10\%$.  
For the $D^+ \to K_{S,L}\pi^+$ case, 
there are more diagrams to consider and predictions take more work.  
CLEO-c results are obtained from 281 pb$^{-1}$ are \cite{CLEO_KSKL}; 
they find $R(D^0) = 0.108 \pm 0.025 \pm 0.024$, 
consistent with $2 \tan^2\theta_C$.  
For the $D^+$, they quote $R(D^+) = 0.022 \pm 0.016 \pm 0.018$.  In this case, 
D.-N.~Gao predicts $R(D^+) = 0.035 \leftrightarrow 0.044$ \cite{gao_rd}, 
while Bhattacharya and Rosner predict 
$R(D^+) = 0.006^{+0.033}_{-0.028} \pm 0.007$ \cite{rosner_rd}.  

We close with a novel result; CLEO-c has measured 
${\cal B}( D_s \to p \bar{n} )
 = (1.30 \pm 0.36^{+0.12}_{-0.16}) \times 10^{-3}$ \cite{cleo_dspn}.  
Well-identified protons are combined with $D$ tags to calculate 
a missing-mass which cleanly peaks at the neutron mass.  
This is the first observation of a charmed meson decaying into a
baryon-antibaryon final state.

\section{Conclusions} 

Many recent experimental charm results have been reviewed, and 
the progress of recent years is continuing.  
Tests of Lattice QCD are becoming more becoming precise, 
and there is currently an intriguing disagreement for $f_{D_s}$ 
even as $f_D$ matches very well.  
Charm threshold data is generally best for experimental precision, 
but significant contributions are also made by other experiments.  
Much existing data is left to mine at BaBar, Belle, and CLEO.  
Very soon we will have data from BESIII and LHC-b, 
and likely a Super-$B$ factory will follow as well.  
In parallel, lattice QCD marches onwards with more CPU and new techniques.  
All in all, charm is alive and well, complimenting and extending $B$ physics.


\begin{acknowledgments}
Thanks are due to my CLEO collaborators P.~Onyisi, M.~Shepherd, T.~Skwarnicki, 
S.~Stone, and W.~Sun, for assistance with this presentation and also to 
my charming BaBar and Belle friends, S.~Prell, Y.~Sakai, and P.~Chang, 
for providing convenient web pages with results.  
I also appreciate the invitation to speak and the efforts 
of the organizers in hosting an excellent meeting.  
\end{acknowledgments}

\bigskip 


\end{document}